\documentclass[prc,twocolumn,showpacs,amsmath,amssymb,floatfix,nofootinbib]{revtex4-1}
\usepackage{graphicx,bm,url,braket,xcolor}

\newcommand{\mybar}[1]{\bar{#1\rule{0pt}{6.5pt}}}
\newcommand{\bb}{$\beta\beta$}
\newcommand{\bbz}{$0\nu\beta\beta$}

\begin{document}

\title{Jastrow functions in double-beta decay}

\author{J. Engel} 
\affiliation{Dept.\ of Physics and Astronomy, University of North Carolina,
Chapel Hill, NC, 27516-3255, USA}

\author{J. Carlson}
\affiliation{Theoretical Division, Los Alamos National Laboratory, Los Alamos,
New Mexico 87545, USA} 

\author{R.B. Wiringa}
\affiliation{Physics Division, Argonne National Laboratory, Argonne, IL 60439}


\begin{abstract}

We use simple analytic considerations and a Monte Carlo calculation of nucleons
in a box to argue that the use of Jastrow functions as short-range correlators
in the commonly employed two-body-cluster approximation causes significant
errors in the matrix elements for double-beta decay.  The Jastrow approach
appears to agree with others, however, if many-body clusters are included.  A
careful treatment of the charge-changing analog of the nuclear pair density
shows, in addition, that differences between Unitary Correlator Operator Method
and Brueckner methods for treating short-range correlations in double-beta are
less significant than suggested by previous work. 
\end{abstract}


\maketitle

\section{\label{sec:intro}Introduction}

New experiments to measure the rate of neutrinoless double-beta (\bbz) decay
will provide information about neutrino masses if neutrinos are Majorana
particles.  Unfortunately, one must know the value of the nuclear matrix
element governing the decay to extract that information from an observed rate
(or rate limit) \cite{avi08}.  For that reason, attempts to better calculate
the matrix elements appear regularly in the literature.  

The matrix elements involve products of one-body decay operators and a sum over
intermediate states, but the closure approximation allows them to be
represented to high accuracy \cite{pan90} by the ground-state-to-ground-state
transition matrix element of a two-body operator 
\begin{equation}
\label{eq:0nuop}
\mathcal{M}^{0\nu} \equiv \sum_{a<b} \mathcal{M}_{ab} \,.  
\end{equation}
The matrix elements $M_{fi} \equiv \bra{f}\mathcal{M}^{0\nu}\ket{i}$ can
therefore be affected by the strong repulsive correlations that alter pair
distribution functions at short distances.  

For many years, theorists were satisfied to simulate these correlations by
using a phenomenological Jastrow function $f(r_{ab})$ in the two-body cluster
approximation to modify the operator $\mathcal{M}$:
\begin{equation}
\label{eq:jastrow}
\mathcal{M}_{ab} \Longrightarrow f(r_{ab}) \mathcal{M}_{ab} f(r_{ab}) \,,
\end{equation}
where $r_{ab}\equiv|\mathbf{r}_a-\mathbf{r}_b|$ is the magnitude of the
distance between the two nucleons.  The Jastrow function almost always had the
form prescribed in Ref.\ \cite{mil76}:
\begin{equation}
\label{eq:miller}
f(r) = 1-e^{-1.1r^2}(1-0.68r^2) \,,
\end{equation}
with $r$ in fm.  Recent work has questioned this prescription.  Refs.\
\cite{kor07,kor07a,kor07b} treated short-range correlations through the
Unitary-Correlation Operator Method (UCOM), which has the advantages of
wave-function overlap preservation and a range of successful applications
\cite{rot05}.  Refs.\ \cite{eng09} and \cite{sim09} computed the effects of
short-range correlations within well-defined Brueckner-based approximation
schemes.  All these papers found smaller effects on matrix elements than the
phenomenological Jastrow function yields.  Because they all limited their
analysis to two-body correlations, however, their predictions for the size of
short-range effects do not come with an iron clad guarantee.  

In fact, all these methods imply the existence of many-body effects that are
always neglected in applications to double-beta decay.  In Jastrow-based
treatments, our subject here, the full correlated wave function is written
schematically as
\begin{align}
\label{eq:corrwf}
\ket{\Psi} = \left(\prod_{a<b<c}T_{abc}\right)\left(\prod_{a<b} F_{ab}\right)
\ket{\Psi_0} \,,
\end{align}
where $\ket{\Psi_0}$ is a Slater determinant or a generalization thereof,
$F_{ab}$ is a two-body Jastrow correlator depending on $r_{ab}$ and the spins
and isospins of particles $a$ and $b$, and $T_{abc}$ is a similar three-body
correlator, which we will ignore from here on.  In recent years, practitioners
have developed a range of techniques for moving beyond the two-body cluster
approximation in Eq.\ (\ref{eq:jastrow}) and including many-body correlations
generated by the product of $F$'s (in addition to explicit three-body
correlations generated by a single $T$) in Eq.\ (\ref{eq:corrwf}).  Cluster
expansions and the Fermi-hypernetted-chain method (see, e.g., Refs.\
\cite{pie92,pan79} and references therein) include three-and-more-body
clusters, and quantum Monte-Carlo methods allow an evaluation of the
contributions of all clusters.  The Jastrow approaches now yield accurate
observables, including two-body density distributions with short-range
correlations, in light nuclei \cite{pie01} and nuclear matter \cite{akm97}.
Here, after analyzing the two-body cluster approximation in \bb-decay, we see
whether an initial application of quantum Monte Carlo with many-body
correlations included supports the phenomenological two-body-cluster Jastrow
method used traditionally, or whether it supports one or more of the approaches
introduced recently.  We also point out that apparent differences between the
results of UCOM and Brueckner methods are largely fictitious.

Heavy nuclei are still too complicated for Monte-Carlo methods in their current
forms, so to evaluate many-body Jastrow effects we look instead at a simplified
version of asymmetric nuclear matter.  We make this choice with the idea that
short-range correlations are nearly universal in nature, depending little on
longer-range structure of the environment in which the correlated nucleons are
embedded, provided that environment has the correct density.  

\section{\label{sec:tbc}Two-body cluster approximation}

\begin{figure}[b]
\includegraphics[angle=0,width=\columnwidth]{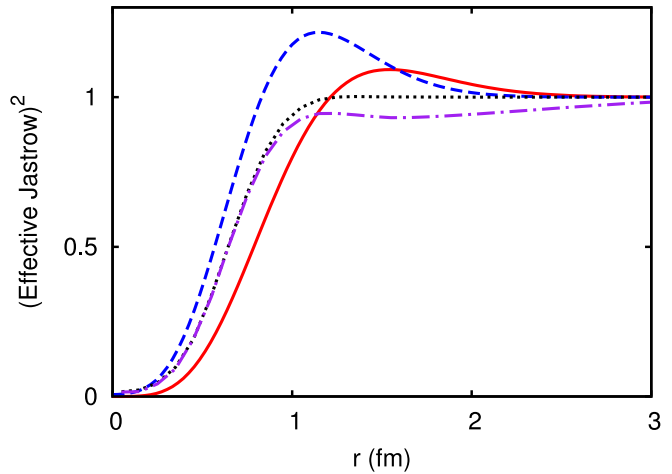}
\caption{\label{fig:1}(Color online) Squares of Jastrow functions $F_{ab}$ from
calculations following Ref.\ \cite{akm97} (dotted black line, spin-singlet
only), from Miller and Spencer \cite{mil76} (solid red line) and from a fit to
the results of a microscopic Brueckner-based calculation \cite{sim09} (dashed
blue line).  The purple dot-dashed line comes from three- and more-body
corrections to the dotted line.}
\end{figure}

In the $S=0$ $T=1$ channel that determines the contribution of short distances
to the \bb\ amplitude, realistic variationally-determined correlation functions
$F_{ab}$ are not so different from the Miller-Spencer Jastrow function.  Figure
\ref{fig:1} shows a typical nuclear-matter example, following the calculations
of Ref.\ \cite{akm97}, alongside the Miller-Spencer function and the effective
scaling function, obtained from the ratio of calculations with and without
short-range correlations, that appears in the Brueckner-based work of Ref.\
\cite{sim09} All the functions go to unity at large $r$, but the
Brueckner-based function has a sizeable "overshoot" near $r=1$ fm.  The
Miller-Spencer function has a much smaller overshoot (occurring at larger $r$,
which is made less important by the radial falloff of the \bbz\ operator)
leading to a significantly smaller \bbz-matrix element.  The variational
nuclear-matter resembles the Miller-Spencer function but has essentially no
overshoot, and so if applied like that function via Eq.\ (\ref{eq:jastrow}) it
will produce an even smaller matrix element.  

The use of the $F$ from Eq.\ (\ref{eq:corrwf}) to multiply a two-body operator
as in Eq.\ (\ref{eq:jastrow}) is often called the two-body cluster
approximation, because all terms are discarded except those in which the
transition operator and the correlators act on the same pair of particles.
This approximation appears to be reasonably good for number-conserving two-body
densities.  The dot-dashed line in Fig.\ \ref{fig:1} displays the distribution
$g_{01}(r)$ in the $S=0,T=1$ channel, following Ref.\ \cite{akm97}, which
incorporates the ful product over all pair correlations of Eq.\
(\ref{eq:corrwf}).  This full $g_{01}(r)$ is somewhat smaller than the
corresponding $F^2$ because many-body tensor correlations promote a fraction of
the spin-singlet pairs to spin-triplet pairs, so that the number of singlet
pairs is reduced. 
The reduction has also been seen in light nuclei \cite{for96}, though the
corrections are not large either there or here.

In \bb\ decay, the picture must be different, however.  To see why, consider
the charge-changing analog of the (spin-independent) two-body density:
\begin{align}
\label{eq:ccdens}
P_F (r) \equiv \bra{f} \sum_{a<b} \delta\left(r-r_{ab}\right) 
\tau^+_a \tau^+_b \ket{i} ~,
\end{align}
where $F$ stands for Fermi.  If we weight this function with $H_F(r)$, the
radial part of the Fermi \bbz\ operator (given approximately by $1/r$), and
integrate, we get the Fermi piece of the \bbz\ matrix element.  If we integrate
$P_F (r)$ without any weighting, we get $\bra{f} \sum_{a<b} \tau^+_a \tau^+_b
\ket{i}$, which must vanish because the isospins of $\ket{i}$ and $\ket{f}$ are
different (in the very good approximation that isospin is conserved), while the
operator between them is proportional to the square of the isospin-raising
operator.

\begin{figure}[ht]
\includegraphics[angle=0,width=\columnwidth]{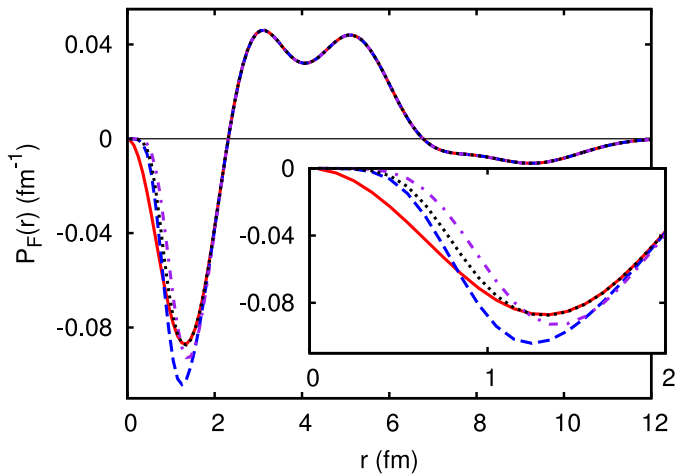}
\caption{\label{fig:2}(Color online) The charge-changing two-body density $P_F
(r)$ for the shell-model calculation of $^{82}$Se in Refs.\cite{cau08,men08}.
The solid red line is the result without short-range correlations, the dashed
blue line is that from the Brueckner-based calculation of Ref.\ \cite{eng09}
the dotted black line applies the Jastrow function from the approach in Ref.\
\cite{akm97} in the two-body-cluster approximation, and the dot-dashed purple
line applies the Miller-Spencer Jastrow function.  The inset magnifies the
left-hand part of the figure.}
\end{figure}

Figure \ref{fig:2} shows $P_F (r)$ for the shell-model calculation of the
\bb-decay of $^{82}$Se in Refs.\ \cite{cau08} and \cite{men08}.  The solid
curve contains no Jastrow function and has area of zero beneath it.  The dashed
curve is the result of of the Brueckner-based calculations in Ref.\
\cite{eng09}.  Its overshoot at $r$ just greater than one causes the integral
to stay very close to zero despite the suppression at very small $r$.  But the
use of the two-body Jastrow function $F_{01}$ \`{a} la Ref.\ \cite{akm97}
(dotted curve) suppresses contributions at small $r$ without an overshoot and
thus leads to an integral of 0.006. Substituting the pair distribution function
$g_{01}$ would only make the problem here worse.  The Miller-Spencer Jastrow
function yields a little bit of overshoot but not nearly enough, and results in
an integral of 0.0075. 

It seems, then, that a realistic treatment of short-range correlations must
yield an overshoot in $P_F (r)$ if it is to preserve isospin (The UCOM
procedure does this exactly, by construction).  When Jastrow functions are
extended beyond the two-body cluster approximation, the effective functions
that result must therefore look different for charge-changing densities, which
involve only valence nucleons, than for like-particle densities, to which all
nucleons contribute coherently.

It is not hard to get an idea of how this happens.  Let us consider
spin-and-isospin-independent two-body correlators $F_{ab}$ (with no three-body
correlators $T_{abc}$) in Eq.\ (\ref{eq:corrwf}) and a general charge-changing
operator $\mathcal{M}_{ab}$.  Writing $F_{ab}^2\equiv 1+h_{ab}$, and expanding
to first order in $h$ gives 
\begin{align}
\label{eq:beyond2b}
\bra{f}\mathcal{M}\ket{i} & = \bra{f_0} \sum_{a<b} \mathcal{M}_{ab}
\prod_{c<d} (1+h_{cd}) \ket{i_0} \\
 &=\bra{f_0} \sum_{a<b} \mathcal{M}_{ab} \ket{i_0} 
 +\bra{f_0} \sum_{a<b} \mathcal{M}_{ab} h_{ab} \ket{i_0} \nonumber\\ 
 &+\bra{f_0} \sum_{\stackrel{a<b}{c\neq \{a,b\}}} \mathcal{M}_{ab} (h_{ac}+ h_{bc}) \ket{i_0} \nonumber \\
 &+ \bra{f_0} \sum_{\stackrel{a<b}{c<d\neq a,b}} \mathcal{M}_{ab}  h_{cd}
 \ket{i_0} + \mathcal{O}(h^2) \nonumber \\
 &= \bra{f_0} \sum_{a<b} \mathcal{M}_{ab} \left(1+\sum_{c<d}  h_{cd} \right)
\ket{i_0} + \mathcal{O}(h^2) \nonumber 
\end{align}
where $\ket{i_0}$ and $\ket{f_0}$ are Slater determinants, and in the third and
fourth lines, $h_{mn}\equiv h_{nm}$ if $n<m$.  

The second line in Eq.\ (\ref{eq:beyond2b}) involves the bare two-body
transition operator and the two-body-cluster correction.  The third line
involves an effective three-body operator, and the fourth line an effective
four-body operator.  Terms of higher-order in $h$ generate even higher
many-body operators. 

Now let the neutron number exceed the proton number so that the Slater
determinants $\ket{i_0}$ and $\ket{f_0}$ have well-defined isospins that differ
from each other, and consider the operator $\mathcal{M}_{ab} = \tau^+_a
\tau^+_b$.  The matrix element above is then the integral of $P_F (r)$, i.e.\
zero.  Although the first term in the second line indeed gives zero, the
second, as we've seen, does not.  The inclusion of all terms first-order in $h$
must restore the value zero, however, because, as the last line shows, the
result can be obtained by acting on $\ket{i_0}$ with the isospin-preserving
two-body operator $\sum_{a<b} h_{ab}$ before acting with the transition
operator.  It is not hard to show that effective four-body term contributes the
same amount as the two-body-cluster correction, and the effective three-body
term contributes twice that amount with the opposite sign, so that the sum of
terms indeed vanishes.  But this also means that, at least to first order in
$h$, three- and-four body effective operators are just as important for the
quantity $\int P_F (r) dr$ as is the effective two-body correction generated by
the two-body cluster approximation.  This perhaps surprising conclusion leads
us to examine the charge-changing density itself and the higher order
corrections in a model amenable to numerical solution. 

\section{\label{sec:nucmat}Many nucleons in a box}

We consider a cubic box with each side $L=4.85$ fm and periodic boundary
conditions.  In the box are 2 protons and 16 neutrons (so that the nucleon
density is very near nuclear-matter density), which decay to 4 protons and 14
neutrons.  The protons in the initial state and all but the last two neutrons
in that state are in filled fermi levels, and the last two (valence) neutrons
are in the spin-zero two-body pairing wave function:  
\begin{align}
\label{eq:boxwf}
\ket{\psi_v} = \mathcal{N} \sum_{k_x,k_y,k_z  \in K} \ket{\mathbf{k},-\mathbf{k};S=0}\,,
\end{align}
where $v$ stands for valence, $\mathcal{N}$ is a normalization constant, and
the set $K$ contains vectors in which two $k$ components are equal to $\pm
2\pi/L$ and the third is zero.  In the final state the neutrons and all but the
last two protons are in filled fermi levels; the two valence protons are in the
configuration $\psi_v$ above, but with the set $K$ containing vectors with one
component equal to $\pm 2\pi/L$ and the other two equal to zero.

\begin{figure}[t]
\includegraphics[angle=0,width=\columnwidth]{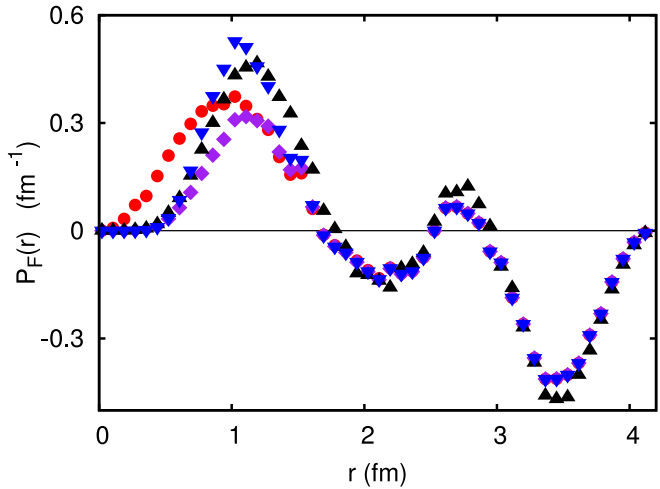}
\caption{\label{fig:3}(Color online) Monte-Carlo calculation of $P_F (r)$ for 2
protons and 16 neutrons in a box decaying to 4 protons and 14 neutrons.  The
red circles are the result with no Jastrow correlators, the purple diamonds
include the Miller-Spencer Jastrow correlator in the two-body-cluster
approximation, the black upward-pointing triangles are the full result with that
correlator, and the blue downward-pointing triangles apply the effective
Brueckner based two-body-cluster Jastrow from Ref.\ \cite{sim09}. The numerical
error associated with the values of $P_F(r)$ are usually smaller than the size
of the corresponding symbols.}
\end{figure}

We use quantum Monte Carlo to evaluate $P_F (r)$ between states of the form
Eq.\ (\ref{eq:corrwf}), where now the states $\ket{i_0}$ and $\ket{f_0}$ are
those just described, and with simple spin-and-isospin two-body correlators
$F_{ab}$ and no three-body correlators $T_{abc}$, as considered previously.
Figure \ref{fig:3} shows the results with the Miller-Spencer Jastrow function.
The two-body-cluster approximation again has very little overshoot, but the
full results, including clusters of all size, has considerably more, so that
the integral vanishes as it should.  Also shown is the result with the
effective Jastrow function fit to the Brueckner-based calculation of Ref.\
\cite{sim09} (which was done in finite nuclei).  It is now quite close to the
full many-body Miller-Spencer result, and the remaining discrepancy is probably
mostly due to the simplicity of our model.  Surprisingly, the use of a Jastrow
function with no overshoot at all gives almost the same result as the
Miller-Spencer function when many-body correlations are taken fully into
account.  

To show the effects of these various functions on \bbz-decay, we define
functions $C_K(r)$, $K$ = $F,GT$, for the Fermi and Gamow-Teller parts of the
\bbz\ operator. (If we wanted to be accurate we would also define one for the
very small tensor term.)  These functions are the products of the densities
$P_F(r)$ and the analogous density 
\begin{align}
\label{eq:ccgtdens}
P_{GT} (r) \equiv \bra{f} \sum_{a<b} \delta\left(r-r_{ab} \right)
\mathbf{\sigma}_a \cdot \mathbf{\sigma}_b \tau^+_a \tau^+_b \ket{i} 
\end{align}
with functions $H_F(r)$ and $H_{GT}(r)$ that specify the radial dependence of
the \bbz\ operators.  In other words
\begin{align}
\label{eq:cdef}
C_K(r) = H_K(r) P_K(r) \,,
\end{align}
with
\begin{align}
\label{eq:hofr}
H_F(r)  &= H_{GT}(r) \approx \frac{2R}{\pi r} \int_0^{\infty} dq \frac{\sin{qr}}
{\rule{0pt}{10pt}q + \mybar{E}-(E_i+E_f)/2}\,.
\end{align}
The quantities $\mybar{E}$, $E_i$, and $E_f$ are energies to which the $H_K$
are not very sensitive.  Equation (\ref{eq:hofr}) holds only if we neglect
nucleon form factors and forbidden currents; in the complete, more complicated
expressions $H_F \neq H_{GT}$ \cite{sim08}.  We use the simplified forms here
because they are sufficient to make our point.

\begin{figure}[ht]
\includegraphics[angle=0,width=\columnwidth]{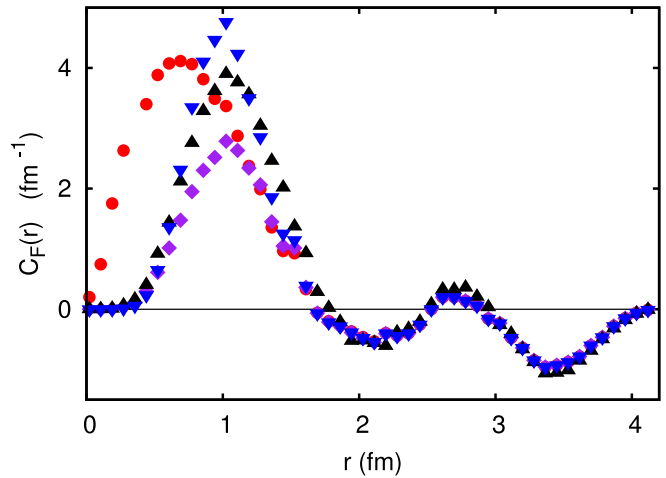}
\caption{\label{fig:4}(Color online) The curves $C_F(r)$ corresponding to the
distributions $P_F(r)$ of Fig.\ \ref{fig:3}, without forbidden currents or
nucleon form factors. The symbols indicate the same approximations as in Fig.\
3.} 
\end{figure}

Figure \ref{fig:4} displays $C_F(r)$ from the Monte Carlo.  (In this simple
calculation $C_{GT}$ is just proportional to $C_F$ because the correlator is
spin-independent and the two valence nucleons that participate in the decay are
locked into a spin-zero configuration).  The full solution clearly corrects the
extreme suppression of the \bbz\ matrix element created by the two-body-cluster
approximation, in a way consistent with the analysis of the integral in section
\ref{sec:tbc}.  Differences with the Brueckner treatment are fairly small and
due once again at least in part to the unusual system we analyze here.  Effects
beyond the two-body cluster approximation are thus both required and apparently
sufficient to describe short-range correlations in \bb\ decay.

\section{\label{sec:ucom}UCOM}

In this section we digress from our main line of inquiry to take up apparent
differences between the results of Brueckner methods and UCOM.  Fig.\ 2 of
\cite{kor07}, Fig.\ 9 of Ref.\ \cite{sim08} and Fig.\ 4 of \cite{sim09} present
\bbz\ distribution functions (similar to the function $C_0^F(r)$ presented here
in Fig.\ \ref{fig:4}) with the UCOM (and other) treatments of short-range
correlations.  Unlike the Brueckner-based curves in our Figs.\ \ref{fig:2} --
\ref{fig:4} the UCOM curves show no overshoot.  But the reason is that the
``contribution from distance $r$'' has been treated differently when UCOM
correlations are considered than when other methods are used.  The UCOM
prescription requires that the operator $r_{ab}$ be replaced by a shifted
version $R_+(r_{ab})$ (where the function $R_+$ is usually determined
variationally) in any operator that doesn't depend on momentum.  Thus, to
evaluate the \bbz\ matrix element, one replaces $H_K(r_{ab})$ by
$H_K(R_+(r_{ab}))$.  

\begin{figure}[b]
\includegraphics[angle=0,width=\columnwidth]{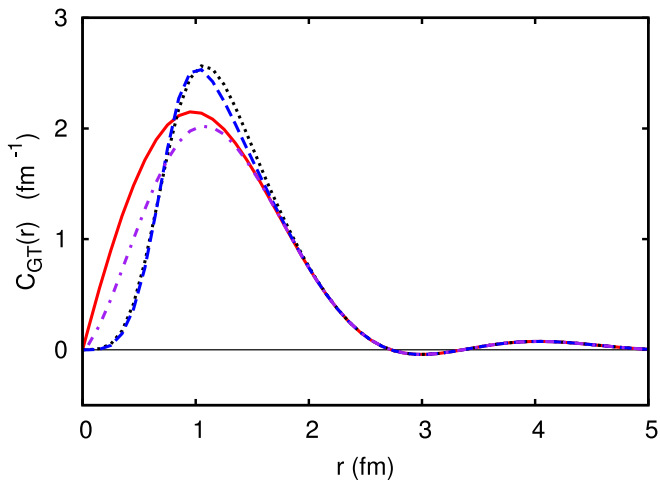}
\caption{\label{fig:5}(Color online) The function $C_{GT}(r)$, defined in Eqs.\
(\ref{eq:ccgtdens}) and ((\ref{eq:cdef}) solid red curve), the corresponding
UCOM function $C^U_{GT}(r)$, defined as in Eq.\ (\ref{eq:newconv}) (dashed blue
curve), the previously-used UCOM function $C^{U\prime}_{GT}(r)$, defined as in
Eq.\ (\ref{eq:oldconv}) (dot-dashed purple curve), and the Brueckner-based
version (dotted black curve).  The new UCOM and Brueckner curves are very
similar.}
\end{figure}

The shifting implies that UCOM produces a function $C_F$ (which we use as an
example because it is simpler than $C_{GT}$) given by
\begin{align}
\label{eq:newconv}
C^U_F(r) & =  \sum_{a<b} \bra{f} H_F(R_+(r_{ab}))  \delta(r-R_+(r_{ab}))
\tau_a^+ \tau_b^+ \ket{i} \nonumber \\
&= H_F(r) \bra{f} \sum_{a<b}  \delta(r-R_+(r_{ab}))
\tau_a^+ \tau_b^+ \ket{i} \nonumber \\
&\equiv H_F(r) P_F^U (r) \,, 
\end{align}
(where $U$ stands for UCOM).  In prior work on UCOM \bb-decay, however, the
``(Fermi) contribution from a given $r$'' was defined instead by simply
replacing $r_{ab}$ with $R_+(r_{ab})$ in $H$, viz: 
\begin{align}
\label{eq:oldconv}
C^{U\prime}_F(r) &= \bra{f} \sum_{a<b} H_F(R_+(r_{ab})) \delta(r-r_{ab}) 
\tau_a^+ \tau_b^+ \ket{i} \nonumber \\
&= H_F(R_+(r)) P_F(r) \,.
\end{align}
This definition, which leaves $r_{ab}$ unshifted in the delta function, gives
the correct result for the Fermi matrix element when $r$ is integrated over,
but does not define an observable and is not what is calculated in other
approaches.  The correct expression, Eq.\ (\ref{eq:newconv}), is a bit more
complicated to evaluate, but has a pronounced overshoot.  Figure \ref{fig:5}
compares the distribution $C_{GT}(r)$ from the shell-model-Brueckner treatment
of $^{82}$Se $\longrightarrow$ $^{82}$Kr in Ref.\ \cite{eng09} (with no
forbidden currents or nucleon form factors) to the properly defined UCOM
distribution for the same decay.  The two curves are extremely close to one
another, and quite different from $C^{U\prime}_{GT}(r)$, which is also shown.
The UCOM and Brueckner pictures are therefore more similar than previously
thought.

\section{\label{sec:disc}Discussion}

The main point of this paper, to which we now return, is that the use of
Jastrow functions in the two-body-cluster approximation suppresses short-range
contributions too much, and that the problem is fixed by including many-body
correlations.  This discovery raises the question of whether existing
treatments are adequate. They include long-range many-body correlations in
shell-model or QRPA wave functions but allow only two particles to be
correlated at short distances.  Is that sufficient? 

It is hard to answer the question definitively because the approach taken here
is so different from the others.  We can say that the very-short-range
correlations are unlikely to be altered; our figs.\ \ref{fig:3} and \ref{fig:4}
show that corrections to the two-body-cluster approximation are barely
noticeable below about $r=0.7$ fm.  But corrections are large at 1 fm or so.
It is far from obvious that the marriage of UCOM or Brueckner treatments of
short-range correlations to shell-model or QRPA treatments of longer-range
correlations incorporates all important effects at the intermediate range $r
\approx 1$ fm.  The UCOM procedure generates three- and higher-body
correlations that have been neglected in almost all applications to date, and
the Brueckner-based double-beta work has so far not included contributions from,
e.g., three-particle ladders.  As for the shell model and QRPA, they leave
untreated a significant range of single-particle energies between those
contained in the calculation and those represented as short-range effects.
Whether these omissions are significant is still an open question.

In the meantime, however, it appears that the UCOM and Brueckner methods give
reasonable short-range correlations.  Like the full Jastrow calculations here,
they supply the overshoot required to preserve isospin symmetry.  Higher-body
corrections in these schemes appear unlikely to be as large as they are in the
Jastrow approach, which violates isospin symmetry in the two-body cluster
approximation.  Short-range effects in \bb-decay thus seem to be mostly under
control.

\begin{acknowledgments}
We gratefully acknowledge the support for this work of the U.S.\ Department of
Energy through the LANL/LDRD Program and through Contract Nos.\
DE-AC52-06NA25396, DE-FG02-97ER41019, and DE-AC02-06CH11357.  Computer time was
made available by Los Alamos Open Supercomputing. 
\end{acknowledgments}

\end{document}